\documentclass[aip,jcp]{revtex4}

\usepackage{graphicx}
\usepackage{tabularx}
\usepackage{color}
\usepackage{amsmath}
\usepackage{mathtools}
\usepackage{comment}
\newcommand{\be}{\begin{equation}}
\newcommand{\ee}{\end{equation}}
\newcommand{\bea}{\begin{eqnarray}}
\newcommand{\eea}{\end{eqnarray}}
\usepackage{dcolumn}
\usepackage{hyperref}
\usepackage{bm}
\usepackage{epsf}
\usepackage{subfigure}
\usepackage{epstopdf}%
\usepackage{amsfonts}
\usepackage{braket}
\usepackage{cancel}
\usepackage{lineno} % DDD
\usepackage{fancyhdr,lipsum}

\begin{document}

\title{From exhaustive simulations to key principles in DNA nanoelectronics}
\author{Roman Korol and Dvira Segal}

\address{
Department of Chemistry and Centre for Quantum Information and Quantum Control, 
University of Toronto, 80 Saint George St., Toronto, Ontario, Canada M5S 3H6
}

%\address{Corresponding author, dvira.segal@utoronto.ca}

\date{\today}

\begin{abstract}
Charge transfer can take place along double helical DNA over
distances as long as 30 nanometers.
However, given the active role of the thermal environment surrounding charge carriers in DNA, 
physical mechanisms driving the transfer process are highly debated.
Moreover, the overall potential of DNA to act as a conducting material in 
nanoelectronic circuits is questionable.
Here, we identify key principles in DNA nanoelectronics by performing an exhaustive computational study.
The electronic structure of double-stranded DNA is described with a coarse-grained model.
The dynamics of the molecular system and its environment is taken into account using a
quantum scattering method,
mimicking incoherent, elastic and inelastic effects. 
By analyzing all possible sequences with 3 to 7 base pairs, 
we identify fundamental principles in DNA nanoelectronics:
The environment crucially influences the electrical conductance of DNA,
and the majority of sequences conduct via a mixed, coherent-incoherent mechanism.
Likewise, the metal-molecule coupling and the gateway states play significant roles 
in the transport behavior.
While most sequences analyzed here are exposed to be rather poor electrical conductors,
we identify exceptional DNA molecules, which we predict to be excellent and robust 
conductors of electric current over a wide range of physical conditions. 
\end{abstract}

\maketitle
%\footnote{*dvira.segal@utoronto.ca}
%================================================

\section{Introduction}
\label{sec-intro}

% Question
Is DNA a good conductor of electricity? 
Experiments evince that
%It is argued that th
a double-helix DNA can support long range charge transfer along its axis % double helix axis
given the good overlap of $\pi$ orbitals of neighboring bases.
Charge transport processes in DNA are central to applications in biology, 
chemistry, physics, and engineering \cite{Tapash}. 
%From the biological point of view, 
Charge transfer plays a crucial role in damage and repair processes in DNA,
thus in the development of cancer in the living cell \cite{Bartonrev10,BartonRev}.
Moreover, DNA is an attractive material for nanoscale engineering and electronic applications:
With its recognition abilities, self-assembly, controllability, structural flexibility, 
and rich electronic properties, % rigidity
it can serve either as the template for nanostructures of desired shapes and function,
or as the conducting compound in molecular electronic circuits \cite{revDNA,bookDNA,BeratanRev,Ron}.

% experiments
Experiments demonstrate highly diverse charge transport behavior through DNA.
In ultra long molecules with hundreds of base-pairs (bp), 
the measured resistance covers the full range of values characteristics to 
metals, semi-conductors, and insulating materials \cite{Danny}. 
Nanoscale (5-20 bp) DNA duplexes are agreed to be conducting, yet measurements demonstrate
a broad range of trends %for the hole transfer rate and the electrical conductance 
\cite{Danny}.
This vast variation is not surprising. %DNA is a complex molecule, and
Charge transfer within a complex molecule such as DNA is influenced by multiple factors.
Depending on the molecular content and the base sequence \cite{Tao04,Arianna,Lewis08,Lewis},
length \cite{Kelly,Giese,Tao04,Slinker, Tao16}, 
backbone composition \cite{Waldeck17},  environmental conditions  \cite{Shao16},
temperature, helical conformation \cite{Hihath15}, linkers to the electrodes
\cite{Harrishole, Waldeck17,Wagner17} and voltage-gating \cite{Taogate},
the four bases of DNA can form sequences that
act, for example, as tunnelling barriers \cite{Tao04,Tao16}, ohmic resistors \cite{Tao04,Tao16,Slinker},
resonant-ballistic or intermediate coherent-incoherent conductors \cite{BartonNN11,Berlininter,Ratner15,Beratan16}. 
It is apparent that exploring charge transport characteristics within a {\it singular} DNA sequence
cannot contribute general insights as to  the potential of the double helix DNA to serve as an electronic material.

%
%Questions
Given this diversity, what valuable information can computational science contribute to DNA electronics? 
Rather than focusing on a particular sequence, we search here for fundamental trends and assess
the capacity of DNA to conduct electric current under different conditions.
The elementary components of DNA are four nucleotides forming
double stranded helix DNA (dsDNA). The strands are hybridized by 
obeying the base-pairing rules, adenine (A) with thymine (T) and cytosine (C) with guanine (G).
Sequences with $3-8$ base pairs are 10-30 \AA  \ long, 
an extended distance for charge transfer in condensed phases.
By studying the electrical conductance of all such DNA sequences,
we henceforth ask the following questions:
(i) What is the distribution of conduction values for these junctions? % and how is it distributed?
%What is the distribution of these values? 
Are most combinations good or poor conductors?
(ii) 
%Working under different environmental conditions, 
What are the physical mechanisms driving charge transfer under different
environmental conditions?  %What are the signatures of these transport mechanisms? 
(iii) How susceptible is DNA conductance to temperature, structural and environmental fluctuations, 
or the contact to the metals?
(iv) Which sequences are excellent conductors---and relatively insensitive to environmental interactions?

% Here

We address these questions with a brute force method, by performing exhaustive numerical simulations.
For a given length (number of base-pairs) $n$, we consider all possible sequences satisfying the base-pairing rules,
and simulate the electrical conductance of these compounds in the geometry of a metal-molecule-metal junction.
Such an ambitious mission can only be accomplished by using a reasonable, economical method.
We employ a tight-binding ladder model to represent the electronic structure of the double helix.
To take into account the impact of structural dynamics (intra and inter-molecular motion), we
perform transport simulations using the Landauer-B\"uttiker's probe (LBP) method 
\cite{Buttiker1,Buttiker2}. %This approach
%allows us to study the electrical conductance of DNA covering coherent, incoherent and in-between  transport mechamisms.
In this approach, the interaction of charge carriers with atomic motion is included 
in a phenomenological manner by introducing a tunable parameter, which is responsible for 
decoherence and energy exchange processes. 
The LBP method covers different transport mechanisms, from pure coherent conduction with frozen nuclei to
the ohmic limit, when electronic coherence is fully lost and diffusive motion prevails
\cite{PastawskiD, PastawskiT,Dhar,Nozaki1,Nozaki2,Qi13,Chen-Ratner,WaldeckF,Hihath15,Kilgour1,Kilgour2,Kilgour3,Korol1,Kim2,Korol2}. 
As well, the LBP method can meaningfully capture intermediate coherent-incoherent transport behavior \cite{Kim1}. 
% conclusions
%

Previous studies were performed on particularly designed, interesting strands, often revealing
%Measurements of charge transfer rates or electrical conductances in dsDNA molecules
signatures of three limiting mechanisms:
(i) quantum coherent ``superexchange" transmission through an A:T block, serving as a tunneling barrier,
(ii) coherent band-like ``molecular wire" conduction, with charge delocalizing along e.g. stacked G:C sequences,
and (iii) sequential multi-site incoherent hopping. 
 %in alternating G:C-rich sequences,
%In the deep tunneling regime, the electrical conductance decreases exponentially
%with distance, therefore becoming negligible in long molecules.
In the latter case,  environmental degrees of freedom such as intramolecular vibrations
localize conducting charges on each G site.
This multi-step hopping process is characterized by an ohmic behavior, a linear enhancement of the 
electronic resistance with molecular length. 
A principal finding of our analysis is that the classification of transport mechanisms as
coherent tunnelling,  coherent-ballistic, or
thermalized multi-site hopping is appropriate for very few strands.
The majority of the examined DNA junctions, operating in ambient conditions, follow an intermediate, 
quantum coherent-incoherent mechanism.
Quantum coherent effects are therefore prevalent and influential
in biological electron transport over distances as long as 40 \AA.

%=======================================================
\section{Model and Method}
\label{sec-model}

\subsection{Setup}
\label{Sexp}
 
We model recent conductance measurements of
relatively short (8-20 bp)  B-form DNA junctions \cite{Tao04,Arianna,Taogate,Ratner15,Tao16,Beratan16}.
This type of experiments are conducted using a scanning tunneling microscope (STM) break-junction approach, 
performed in aqueous solution or in humidified atmosphere at room temperature. 
In a typical experiment, both the STM tip and the substrate are made of gold,
but the tip is further coated with a wax layer to minimize ionic conduction between the electrodes.
To ensure strong, chemical binding of the molecule to the electrodes,
each DNA molecule is modified with thiol or  amine linkers. 
The substrate is immersed into  a %phosphate 
buffer solution containing dsDNA. It is
then dried with a nitrogen gas, with measurements performed in a humidified atmosphere.
STM break-junction measurements are performed thousands of time by repeatedly bringing the tip into and out of 
contact with the substrate---which is covered with dsDNA molecules within a water layer. 
The electrical conductance of  the junction created in the pulling processes is recorded as a function of the 
tip-substrate distance; a plateau in the retracting curve indicates on the formation of a molecular junction. 
Conductance histograms provide the most probable conductance value, as well as estimates over the number of molecules
forming the junction and the heterogeneity of the contact geometry.

%--------------------------------

\subsection{Electronic Hamiltonian}
\label{Sele}

\begin{figure}[htbp]
\hspace{-9mm}
%\vspace{-10mm}
%\includegraphics[width=2.5cm]{s.eps}
\includegraphics[width=17cm]{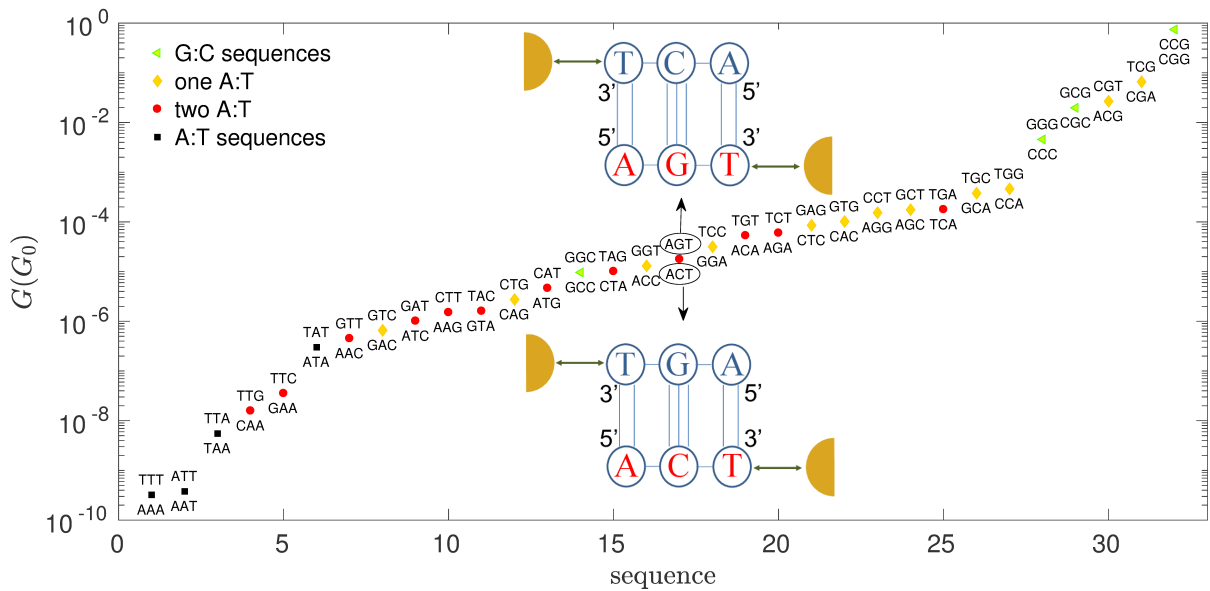}  % 3bp4
\caption{Electrical conductances of all 32 combinations of 3 bp DNA junctions, 
$\gamma_d=0$, $\gamma_{L,R}=50$ meV,  $T_{el}=5$ K.
%For $n$=3, there are $4^3$ = 64 sequences with 32 different pairs.
By convention, sequences are labeled from the 5' to the 3' end.
We further sketch the geometry of examined junctions, with the DNA molecule connected to the metals (semi-spheres) at its 3' ends.
We assume that the metal contacts are identical, thus,
given symmetry, the two sequences displayed show an indistinguishable conductance.}
\label{3bp}
\end{figure}

For a given number of base pairs, we draw all possible DNA duplexes; there are $4^n$ such molecules
composed from the four nucleotide bases.
Molecules are connected to the metals through the 3' ends, see Figure \ref{3bp}, mimicking experiments.
We model the electronic properties of B-form DNA using a coarse-grained tight-binding ladder Hamiltonian,
see e.g. Refs. \cite{cunibertiphonon, Rydnyk,ladder1,ladder2,ladder3,BerlinJacs,Wolf}.
There is an extensive evidence that holes, rather than electrons, dominate charge migration in DNA, and
that charge transport takes place inside the double helix along the  $\pi$ stacking,
rather than through the sugar-phosphate skeleton \cite{Kurita,Harris,Yoshizawa}.
The electronic Hamiltonian describing hole migration in a dsDNA junction 
reads $\hat H=\hat H_M+\hat H_L+\hat H_R + \hat V_L+\hat V_R$. The molecular term is
\bea
\hat H_M&=&\sum_{j=1}^n \Bigg[
\sum_{s=1,2} \epsilon_{j,s}\hat c_{j,s}^{\dagger}\hat c_{j,s} % energies
+ \sum_{s\neq s'=1,2} t_{j,ss'}\hat c_{j,s}^{\dagger}\hat c_{j,s'}
\nonumber\\
&+&
 \sum_{s,s'=1,2} t_{j,j+1,ss'}(\hat c_{j,s}^{\dagger}\hat c_{j+1,s'} + h.c.) \Bigg].
\label{eq:HM}
\eea
%
%This Hamiltonian describes an $n$ base-pair long  dsDNA molecule. 
Each site represents a particular base, $N=2n$ is the total number of bases.
The index $s=1,2$ identifies the strand. %. , see Fig. XXX to see our convention.
$\hat c_{j,s}^{\dagger}$ creates a hole on strand $s$
at site  $J$ with an on-site energy $\epsilon_{j,s}$. $t_{j,ss'}$ and $t_{j,j+1,ss'}$
are the electronic coupling elements between nearest neighboring bases.
The model mimics the topology of dsDNA molecules; helical effects are
taken into account within renormalized electronic parameters.
The electrodes ($L$,$R$) are modeled as Fermi seas of noninteracting electrons with 
$k$ as the index for momentum,
(fermionic creation operators $\hat a_{k,L/R}^{\dagger}$),
\bea
\hat H_{L}&=&\sum_{k}\epsilon_{k,L}\hat a_{k,L}^{\dagger}\hat a_{k,L},\,\,\,\ 
\hat H_R=\sum_{k}\epsilon_{k,R}\hat a_{k,R}^{\dagger}\hat a_{k,R} .
\eea
The first (last) site on the $s=1$ ($s=2$) strand is coupled to the left (right) metal lead, 
%the last site on the $s=2$ strand is coupled to the right electrode,
%
\bea
\hat V_{L}=
\sum_{k}g_{k,L}\hat a_{k,L}^{\dagger}\hat c_{j=1,s=1} + h.c. ,\,\,\,\,\
\hat V_{R}=
\sum_{k}g_{k,R}\hat a_{k,R}^{\dagger}\hat c_{j=n,s=2} + h.c.
\eea
%
% param
We adapt a DFT- based parametrization;
electronic site energies and  matrix elements, $t_{j,ss'}$ and  $t_{j,j+1,ss'}$,  are listed in Ref. \cite{BerlinJacs}. 
To introduce relevant energy scales, in Table \ref{table:en} we list site energies, reported relative to the guanine base,
and inter-strand coupling \cite{Consite}.

%================================================================================================
% Table I
\begin{table}[ht]
\begin{tabularx}{.45\textwidth} { c c  c  c c c  }
\hline
\hline
%$\epsilon_G$ \hspace{3mm} & $\epsilon_A$ \hspace{3mm} & $\epsilon_C$ \hspace{3mm} & $\epsilon_T$ \hspace{3mm} & $t_{\rm{G||C}}$ \hspace{3mm} &  $t_{\rm{A||T}}$\\
$\epsilon_G$ \hspace{3mm} & $\epsilon_A$ \hspace{3mm} & $\epsilon_C$ \hspace{3mm} & $\epsilon_T$ \hspace{3mm} & $t_{\rm{G:C}}$ \hspace{3mm} &  $t_{\rm{A:T}}$\\
\hline
%8.178 \hspace{3mm} & 8.631 \hspace{3mm} & 9.722 \hspace{3mm}& 9.464 \hspace{3mm} & -0.055 \hspace{3mm}& -0.047 \\
 0 \hspace{3mm} & 0.453 \hspace{3mm} & 1.544 \hspace{3mm}& 1.286 \hspace{3mm} & -0.055 \hspace{3mm}& -0.047 \\

\hline
\end{tabularx}
\caption{On-site energies, relative to the guanine base, and inter-strand electronic coupling $t_{j,s\neq s'}$ (eV).
The full parameter set for the Hamiltonian (\ref{eq:HM}) is included in Ref. \cite{BerlinJacs}.}
\label{table:en}
\end{table}

%===========================================================================================

%LBP
%\subsection{Condensed phases environment} %: Landauer-B\"uttiker probe method}
\subsection{Nuclear Environment} 
\label{Senv}

Charge transport through DNA is critically influenced by the surrounding thermal environment, comprising
nuclear dynamics of the nucleobase, structural motion, reorganization of solvent molecules around the transferred
charge, polarization effects through the backbone and counterions.
This fluctuating and correlated environment can be captured, for example,
with a coarse-graining approach, by building the effect of the environment on the electronic Hamiltonian
within spatially and temporally corrected noise terms \cite{flickering,BeratanZhang}.
Other approaches are based on combining classical molecular dynamics (MD) simulations
with quantum mechanics/molecular mechanics (QM/MM) methodologies
% The coarse-grained electronic Hamiltonian is constructed along the MD trajectory 
%used to compute the transmission function or the charge transfer rate
%from the MD trajectory 
\cite{cunibertiJCP09,cunibertiLee,cunibertiNJP10,kubar1,kubar2,kubar3}.
One may also explicitly consider the interaction of transport charges with
selected internal vibrational modes using the Green's function formalism \cite{cunibertiphonon},
quantum rate equations \cite{PeskinPCCP}, or semiclassical approximations \cite{BerlinSC}.

Here, we use an alternative, low-cost technique and account for system-environment interactions by employing 
the Landauer-B\"uttiker probe method \cite{Buttiker1, Buttiker2}. 
It is applicable for the study of charge conduction in 
a wide range of systems, from single-atom point contacts up to the thermodynamic limit
\cite{PastawskiD, PastawskiT,Dhar,Nozaki1,Nozaki2,Qi13,Chen-Ratner,WaldeckF,Hihath15,Kilgour1,Kilgour2,Kilgour3,Korol1,Kim1,Kim2,Korol2,CasatiRev}.
In this technique, incoherent elastic and inelastic electron (or hole) 
scattering effects are taken into account  %in a phenomenological manner
by augmenting the non-interacting electronic Hamiltonian with probe terminals through which 
charge carriers loose their phase memory
and possibly exchange energy with other degrees of freedom.
The technique was originally introduced to study decoherence effects in mesoscopic devices,
yet it was successfully applied to explore electronic conduction in organic and biomolecular systems 
\cite{Nozaki1,Nozaki2,Chen-Ratner,WaldeckF,Hihath15}.
%anharmonic effects in (purely) phononic quantum conduction \cite{SC2,SCMalay,SCnano}.

% our work on LBP
We had recently demonstrated that the LBP method can capture different,
limiting transport regimes in molecular transport junctions: tunneling conduction,
ballistic motion, and incoherent hopping \cite{Kilgour1}. 
Moreover, the method  can reproduce an intermediate quantum coherent-incoherent transport behavior
in a qualitative agreement with experiments on DNA junctions \cite{Kim1}. 
As described in e.g. Refs. \cite{Salil,Kilgour3}, 
the LBP method can be applied in different fashions so as to control scattering events.
Here, we use the so-called ``voltage probe" method at low bias, 
which implements elastic and inelastic scattering processes under low applied bias---within linear response.

The methodology was detailed elsewhere \cite{Kilgour1,Kilgour2,Kilgour3,Kim1,Kim2,Korol1} and the program
was published in \cite{Korol2}. 
Here, we recount only the essential principles so as to introduce  working parameters.
We voltage bias the electrodes, $\Delta \mu =\mu_L-\mu_R=eV$
and fix the temperature of the metals at $T_{el}$. %and the environment, the latter corresponds to nuclear degrees of freedom.
Each electronic site (base), 
is connected to a ``probe", emulating the dynamical environment, with the hybridization energy $\gamma_d$.
Charges are allowed to leave the molecule towards the probes, 
where they loose phase information and absorb or release energy. 
Nevertheless, we set the chemical potentials of the probes such that 
the current in the physical system is conserved. 
The electrical conductance of the junction, in units of $G_0=e^2/h$, with the electron charge $e$ and Planck's constant $h$, is given by
\bea
G=\frac{1}{V}\sum_{\alpha} (\mu_L-\mu_\alpha) \int_{-\infty}^{\infty}\mathcal T_{L,\alpha}(\epsilon) 
 \left(-\frac{\partial f_{eq}}{\partial \epsilon}\right) d\epsilon.
%\left[f_L(\epsilon)-f_{\alpha}(\epsilon)\right]d\epsilon.
\label{eq:currL}
\eea
The summation over $\alpha$ includes the physical, source and drain electrodes as well as the probes.
Here, $f_{eq}(\epsilon)=[e^{\beta_{el}(\epsilon-\epsilon_F)}+1]^{-1}$ is the equilibrium Fermi-Dirac
distribution function, given in terms of the
temperatures $k_BT_{el}=\beta_{el}^{-1}$ and the Fermi energy $\epsilon_F$.
The transmission function in Eqs. (\ref{eq:currL}), 
$ \mathcal T_{\alpha,\alpha'}(\epsilon)={\rm Tr}[\hat {\Gamma}_{\alpha'}(\epsilon)
\hat G_{r}(\epsilon)\hat \Gamma_{\alpha}(\epsilon)\hat G_a(\epsilon)]$, is
expressed in terms of %an ($N\times N$) sized 
the Green's function, 
$G_{r}(\epsilon)=1/(\epsilon\hat I-\hat H_M-\hat \Sigma)$;
$\hat G_a(\epsilon)=G_r^{\dagger}(\epsilon)$, and $\hat I$ as the identity matrix.  %The trace is performed over the electronic states of the molecule. 
The self energy $\hat \Sigma$ includes contributions from the electrodes, 
with the hybridization 
matrices $\hat \Gamma_{\alpha}(\epsilon)=2{\rm Im} \hat \Sigma_{\alpha}(\epsilon)$ \cite{nitzan,diventra}. 
Assuming energy independent functions and local couplings,
these matrices include a single nonzero element \cite{gamma}.
%%
%We count the bases $i=\{1,2,...n, n+1,n+2,...,N\}$, listing first sites from the $s=1$ strand, then the second one.
The metal-molecules hybridization energy is given by $\gamma_{L,R}$. 
Environmental effects are captured by the parameter $\gamma_d$, with $\hbar/\gamma_d$ 
as the characteristic elastic and inelastic scattering time of charge carriers.
Lastly, to evaluate Eq. (\ref{eq:currL}), 
the chemical potentials of the probe terminals are determined from an 
algebraic charge-conservation equation \cite{Kilgour1,Kilgour2,Kilgour3,Korol1}.
One should note that the probes not only introduce level broadening, but further 
open up additional, incoherent transfer pathways, as indicated by 
the appearance of new transmission functions in Eq. (\ref{eq:currL}), beyond the direct $L$ to $R$ contribution.

%=============================================

%---------------
% Figure 1
%---------------

\section{Results and Discussion}
\label{sec-simul}

\subsection{Choice of parameters}
\label{Smech}

%=================================

Our model comprises a parameterized electronic tight-binding Hamiltonian for the DNA duplex
and additional parameters, %for characterizing the junction and its dynamical environment: 
$\gamma_d$, capturing  scattering effects of carriers due to molecular  dynamics,
and $\epsilon_F$, $T_{el}$  and $\gamma_{L,R}$,
describing the electrodes and the metal-molecule hybridization.
We now explain our choice of simulated values.
%
% gd
The probes emulate decoherence and energy exchange processes 
at a rate $\gamma_d/\hbar$. % a tunable parameter controlling system-environment interactions.
For DNA in aqueous or humidified conditions at room temperature,
previous simulations suggested %% temp?
$\gamma_d=5-30$ meV \cite{Qi13,Hihath15,Kim1,Kim2, Korol1}.
We further justify this  range  as follows.
First, calculations corresponding to dry DNA yield $\gamma_d=$  1-6 meV \cite{Qi13}, but
charge transfer within a wet medium is expected to suffer from stronger environmental effects.
Moreover, molecular dynamics simulations of DNA in solution  % XXX
suggest that fluctuations of site energies have a lifetime  $\tau\sim $200 fs
\cite{voityuk,BeratanZhang}, which converts to $\hbar/\tau=$ 20 meV,
within the range of our estimated decay constant.

We perform simulations  at $\gamma_d=0$, corresponding to charge transfer in rigid-frozen structures,
and at $\gamma_d=$ 10 and  30 meV,
representing dsDNA in solution at ambient conditions.
We further run simulations at lower ($1$ meV) and higher ($50$ meV) values,
and confirm that observed phenomena are
regularly-monotonously modified by this parameter.
Note that electronic tunneling energies in DNA  are order of 1-75 meV \cite{BerlinJacs}.
We do not  explicitly introduce a temperature for the nuclear (molecular, solvent) degrees of freedom, 
as this temperature is encoded in the magnitude of $\gamma_d$.
Temperature, denoted by $T_{el}$, explicitly appears in
Eq. (\ref{eq:currL}) and it dictates electronic population in the metals,
thus the broadening of the Fermi function.
At high electronic temperature, carriers fill the tail of the Fermi function, which  is
in resonance with molecular orbitals. 
This contribution is reflected by an enhanced resonant-ballistic current.
In real systems, this delicate, resonant contribution is quickly suppressed by temporal fluctuations of the structure \cite{flickering}.
To capture this suppression effect and reduce the impact of the resonant-band like current,
we further test our simulations at a rather low electronic temperature,
$T_{el}= 5$ K, bounding injected electrons to the vicinity of the Fermi energy.

We employ two representative values for the metal-molecule
coupling, $\gamma_{L,R}$=50, 1000 meV, corresponding to moderate and strong metal-molecule hybridization.
Another tunable parameter is the position of the Fermi energy of the metals 
relative to molecular states.
Since the HOMO level of DNA appears close to the Fermi level of gold,
compared to its LUMO level, holes (rather than electrons) are the main charge carriers in DNA \cite{Giese}. 
The measurement of a positive Seebeck coefficient  \cite{Tao16}  further affirms this conclusion. %that hole conduction dominates in DNA.
Specifically, since the HOMO of the guanine nucleotide lies close to the Fermi energy of the gold electrode,
we place $\epsilon_F$ on resonance with the on-site energy of the guanine base
%the Fermi energy of the metals in resonance with the on-site energy of the guanine base, see Table I
 \cite{Rydnyk,Qi13,Hihath15,Kim1,Kim2,Korol1}. % CHECK

%NEW
We demonstrate below the behavior of  $n=3-7$  long sequences, but we had further looked at 
all  combinations with $n=8$ base pairs. Our observations are retained throughout.
Given the strong signatures of observed effects, we expect our conclusions to certainly hold for molecules 
with $n=10$ base pair.

%---------------
%Figure 2
% G5_Gd_T5.fig and figdeph5
\begin{figure}[htbp]
%\hspace{-3cm}
\includegraphics[width=19cm]{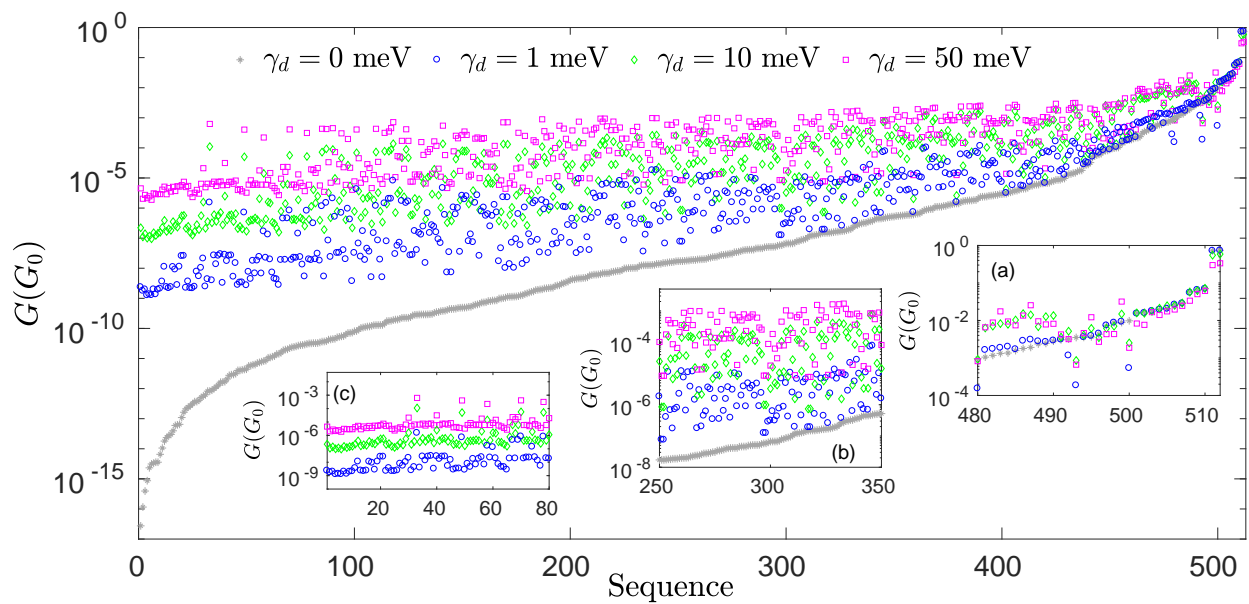}                   %{figdeph5.eps} G5a
\caption{Conductances of dsDNA with 5 base pairs under increasing environmental effects,
$\gamma_d=0$, 1, 10 and 50 meV.
Other parameters are  $\gamma_{L,R}=50$ meV and $T_{el}=5$ K.
}
\label{5bpgd}
\end{figure}
%---------------

\subsection{Characteristic trends}

To illustrate our results,  Figure \ref{3bp}  presents the conductance of short ($n=3$), rigid junctions. 
Molecules are connected to the electrodes at the 3' terminal, 
and sequences are labeled by convention 5' to 3'.
Every point in this graph corresponds to two
sequences, which due to symmetry yield identical values.
For an odd number $n$, there are $4^n/2$ different sequences.
%Therefore, there are 32 distinct junctions for $n=3$.
For an even $n$, there are $4^{n/2}$ palindrome sequences such as 5'AATT3',
which do not have a twin. Therefore, for $n$=4 we identify $(4^n-4^{n/2})/2+4^{n/2}$=136 distinct junctions.
Counting in this manner, there are 32, 136, 512, 2080, 8192, and 32,896 distinct junctions for $n=3,4,5,6,7,8$, respectively.
%2080 distinct junctions for $n=6$, and
%$4^7/2=8192$ different junctions for $n=7$.

Sequences in Figure \ref{3bp} are presented in order of an increasing conductance. 
The color scheme emphasizes general trends: Sequences rich in G:C base pairs are good conductors, 
while A:T-rich sequences are poor conductors
(recall that the Fermi energy is placed at the energy of the G base).
Notably, the sequence GCC is not an impressive conductor given that the C bases are placed at the edges.
This demonstrates that beyond composition, gateway states
allowing for a forceful injection of charge, are crucial for organizing an excellent conductance.
Overall, the conductance of 3 bp DNA covers almost 10 orders of magnitude.
The best conductor in Fig. \ref{3bp}, CGG, has a non-interrupted path of G-bases, with a single crossing 
between strands. 
Next in order are sequences with a single adenine base that carriers need to traverse. 
Sequences GGC and CCC are significantly lower in conductance
even though they contain only G:C base pairs,
since the highest-energy C base is linked to one of the terminals.
Finally, sequences with an A:T block are rather poor conductors.

We now include environmental effects,
% for structural motion---environmental effects---
captured by a nonzero probe coupling $\gamma_d$. 
For clarity, Figure \ref{5bpgd} depicts the conductance of an $n=5$  long molecule, but
we confirmed that our results are representative for longer chains, $n=6-8$.
We organize the sequences in order of increasing
conductance at $\gamma_d=0$, and find roughly three families of rigid molecules: good, poor, and intermediate 
conductors. Upon turning on the environment, we identify the following trends:

(i) Good conductors with $G=0.01-1$ $G_0$ are only mildly affected by the environment, see panel a.
This indicates that in this case charges proceed through delocalized molecular states.
If fact, the conductance here is slightly reduced with the increase of system-bath coupling
since scattering processes hamper delocalized motion.

%\item 
(ii) Poor rigid conductors, 
sequences 1-100 with $G\lesssim10^{-10}$ $G_0$, 
enjoy a dramatic enhancement of their conductance due the environment. %, by 5-10 orders of magnitude. 
This giant increase over 5-10 orders of magnitude 
indicates that the underlying transport mechanism has been changed, most likely
from off-resonant tunnelling to environmentally-assisted transport.
The observed  scaling with $\gamma_d$ 
further evinces that charge carriers proceed through multi step hopping 
\cite{Kilgour1}. A curious observation is that 
under this mechanism the conductance of sequences 1-100 is almost a constant, see panel c,
%It is also interesting to note that in the hopping limit (sequences 1-100), the conductance is about 
insensitive to composition and order, which is in a stark contrast to rigid molecules,
with sequences 1-100 showing five orders of magnitude variation in conductance.

(iii) Intermediate conductors ($G=10^{-6}-10^{-2}$ $G_0$) display significant variability under environmental effects.
Sequences that similarly conduct when rigid, greatly digress once environmental effects take place, see panel b.
%aAs we show below, %in Figs. \ref{7bpgd} and \ref{7bpglr}, 
This spreading demonstrates that
one should not assume that the quantum coherent (frozen) value represents in any way 
the behavior of a flexible system. Moreover, these intermediate sequences cannot be categorized
 as tunneling/ohmic/ballistic conductors.
Finally, as we show below, this variability corresponds to 
clustering of A:T units vs. spreading them apart, which notably only affects the conductance of non-rigid structures.
%\end{itemize}

%---------------
% Figure 3
% histogram at $gamma_d=0$
\begin{figure}[htbp]
\includegraphics[width=16cm]{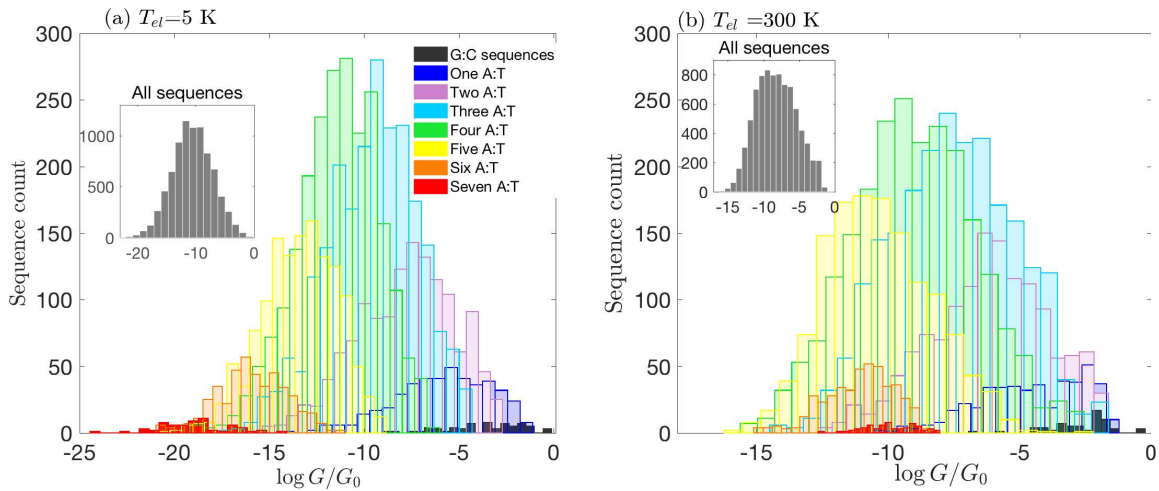} % figure4TT
\caption{Conductance histogram for rigid $n=7$ bp DNA, separated into
sequences with different number of A:T base pairs. The inset presents data for all combinations.
(a) $T_{el}=5$ K, (b) $T_{el}$=300 K. Other parameters are
$\gamma_{L,R}=50$ meV and $\gamma_d$=0 meV.}
\label{7bphist1}
\end{figure}
%---------------

\subsection{Composition} %: Number of A:T base pairs in the sequence}

Which DNA sequences are poor electrical conductors?
Let us first focus on rigid molecules,
% in Figure \ref{7bphist1},
and watch in Figure \ref{7bphist1}  the  distribution of conductances for an $n=7$ bp DNA, 
as well as histograms for different compositions.
%at $\gamma_d=0$. 
We find that the rule of thumb, G:C rich sequences being good conductors, is valid.
Moreover, at low electronic temperature A:T sequences act as rather poor electrical conductors.
One should note however that at room-temperature,
sequences with a stacked A segment can support ballistic, band-like current, 
thus they may conduct more effectively than mixed-nucleotide sequences.
Interestingly, the histograms resemble a normal (Gaussian) distribution.
The mean of the histogram corresponds to a uniform sequence with an
averaged barrier height. The width signifies a strong sensitivity to the 
arrangement of base pairs within the sequence.

In table \ref{table:seq7}, we exemplify
sequences within the three groups: insulators, weak-to-moderate conductors,
and excellent conductors. %Sequences are listed from the 5' to the 3' ends.
Poor conductors include an A:T block. 
Having a C base as the edge results in poor conduction.
%$\epsilon_C-\epsilon_G~ 1.7$ eV, is the highest energy gap.
Excellent conductors  allow delocalization of charges throughout the sequence.
For example, in the %CCCCCCG 
CGGGGGG sequence, charges enter through the G base, cross to the other strand only once,
and continue through the G block un-interrupted until the other metal. 

%===============
 %Sequences are listed from the 5' to the 3' ends.
% TABLE II
\begin{table}[ht]
\caption{Examples for poor, moderate and excellent  $n=7$ conductors. 
Conductance is reported for a frozen molecule, $\gamma_d=0$ at $T_{el}=5$ K.
 }
\label{table:seq7}
\begin{tabular}{|l|l|l|l|}
 \hline
Poor,  $G<10^{-11}$   $G_0$   &  weak-moderate ,  $10^{-8}<G<10^{-3}$  $G_0$ &  good-excellent, $G>10^{-2}$ $G_0$     \\
 &    &        \\
\hline
   ATTTAAA (10$^{-21}$) &     ATGGGCA  (5 $\times $10$^{-7}$) &   CCCCGGA (0.010) \\   % 
  AAAAAAA  ( $10^{-20}$) & CGCATGC ($10^{-6}$) & CACCCGG(0.015)     \\   %
   AATGAAA ($10^{-18}$)&     ACGATGG($10^{-6}$) &   CCCCGAG (0.060) \\   %
   CCTAAAA ($10^{-16}$)&  CTCGCGA ($10^{-5}$)&  CGGGGGA (0.064) \\    % 
    CAACAAT ($10^{-15}$)&  ACGCGGC ($10^{-5}$)    &   CCGGGGG (0.70) \\     % 
  AAGAATA ($10^{-14}$) &     CTCCGTG ($10^{-4}$)   &CGGGGGG  (0.75) \\ %
 \hline
\end{tabular}
\end{table}
%=====================

 %---------------
% Figure 4
\begin{figure}[htbp]
\includegraphics[width=18cm]{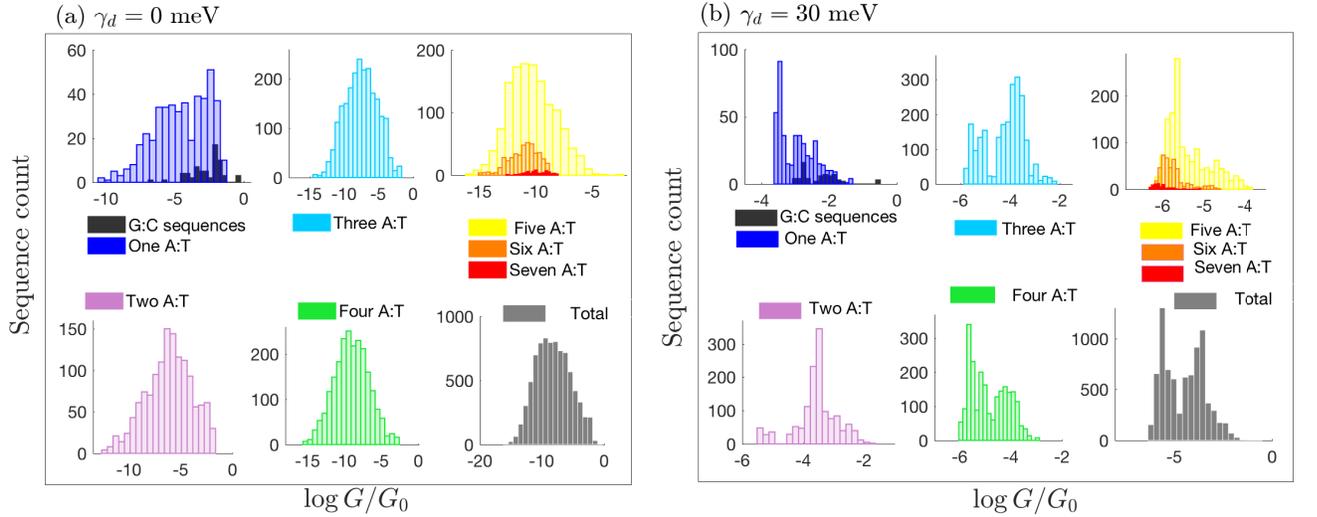} % histSSSS
\caption{Histograms of conductance for $n=7$ bp DNA, divided into
groups with different content.
(a) $\gamma_d=0$, (b) $\gamma_d=30$ meV. Other parameters are
$\gamma_{L,R}=50$ meV and $T_{el}$=300 K.  The
insets present conductance histogram for all 8192 sequences.}
\label{7bphist2}
\end{figure}
%---------------

\subsection{ Environmental effects} %: $\gamma_d\neq0$}

We turn on the environment in Figure \ref{7bphist2},
and expose its striking role on the conductance  histograms.
We find that distributions 
at finite $\gamma_d$ not only shift to higher values, compared to the frozen case, %since multi-step hopping conduction contributes,
but further {\it split}, showing up as bimodal distributions with a gap in the middle. % (see \ref{Fig7})
In Table \ref{table:struc} we list several sequences that support a comparable 
electrical conductance when frozen,
but receive values 1-2 orders of magnitude apart at finite system-environment interaction.
The opening of the gap in the distribution can be rationalized:
In rigid structures, the averaged barrier height is a principal variable, as it controls off-resonant conduction.
In contrast, once interacting with the environment, carriers can partially localize on the $G$ bases,
and hop between them.  In this scenario, an A:T segment, such as  in AAATCGG,
significantly hampers the overall conductance compared to the case with isolated A:T units, 
as in ACAGTGT, see Table \ref{table:struc}.
The two components in the bimodal distribution therefore correspond to sequences with clustered vs.
desolate A:T base pairs. 
We thus arrive at a critical observation: In rigid structures,
the composition essentially determines the conductance, 
but the structure (base order) is of a lesser importance.
In contrast,  
in flexible molecules  transferred charges
are highly sensitive to the development of local, enlarged barriers, 
which disturb site-to-site hopping dynamics.

%===============
% TABLE III
\begin{table}[h]
\caption{
%$\log_{10} G$ in units of $G_0$ for 
Role of the structure (clustering) on DNA conductance,
$T_{el}=300 $ K, $\gamma_{L,R}$=50 meV}.
\label{table:struc}
\begin{tabular}{|c|c|c|c|c|c|}
 \hline
 &    &    &   &    & \\
Sequence  & \,\ $\#$ of A:T bp \,\ & \,\ $\log G/G_0$ \,\  &  \,\ $\log G/G_0$ \,\ &\,\  $\log G/G_0$ \,\     &  \,\ A:T block? \,\ \\  % log 10
5' to 3'& & $(\gamma_d=0$ meV)  &   $(\gamma_d=10$ meV) &    $(\gamma_d=30$ meV)    &   \\
 &    &    &   &    & \\
\hline
% &    &    &   &  &  \\
   AATGCGC  & 3 & -8.0  &   -6.0  &  -5.3   & yes\\   % 
   ACAGTCG &  3 &-8.0  &    -4.1 & -3.7  & no \\
% &    &    &    & & \\
\hline
% &    &    &   &   & \\
  AAATCGG & 4&-8.0 & -6.5  & -5.7 & yes \\
  ACAGTGT & 4&-8.0 &  -3.8  &  -3.4 & no\\
% &    &    &   &    &\\
\hline
 %&    &    &   & & \\
 AAAAAGG &5 &-11.6 & -7.0 &  -6.1& yes \\
% AACTAAC  &5 & -11.5 & -5.6 &  -5.0 &  \\ 
 AACAGTA & 5 &-11.6 &  -4.3 &  -4.1  & no\\
%&  &    &    &   & \\
 \hline
\end{tabular}
\end{table}
%=====================

% Figure 5
%---------------
% G7_G_Gd0_1.fig
% scatter1a in the original place
\begin{figure}[h]
\vspace{-4mm}
\includegraphics[width=14cm]{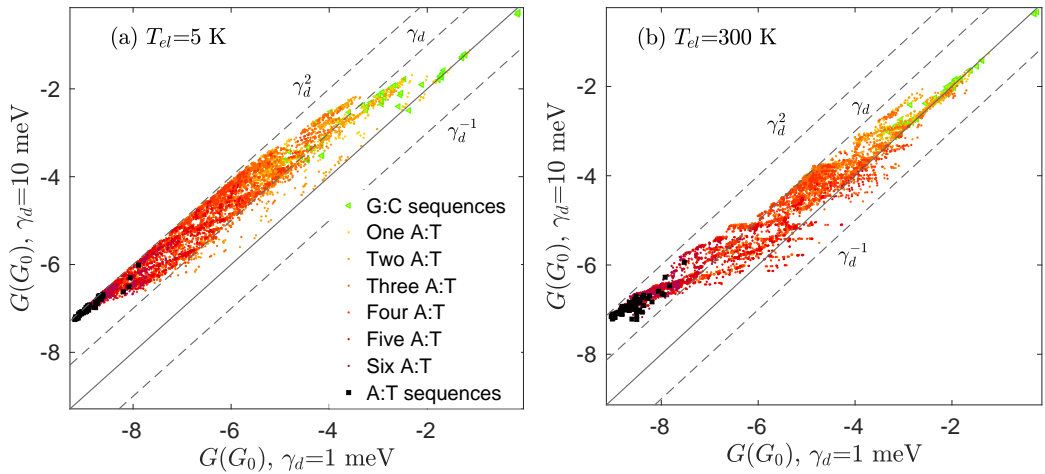} % scatter1h
\caption{Effect of environmental fluctuations on the conductance of all $n=7$ bp DNA molecules, 
$\gamma_{L,R}=50$ meV, (a) $T_{el}=5$ K, (b) $T_{el}=300$ K.
The diagonal (full) identifies sequences that are undisturbed by the environment. 
The scaling $G\propto \gamma_d^{-1}$, $G\propto \gamma_d$, and $G\propto \gamma_d^2$ are marked by dashed lines.}
\label{7bpgd}
\end{figure}
%---------------

What is the principal physical mechanism driving charge transfer in dsDNA? 
In Figure \ref{7bpgd} we display the conductances of $n=7$ bp DNA
for two different values of $\gamma_d$, 1 meV and 10 meV. 
Each dot corresponds to a particular sequence, and we make the following observations:
(i) The best conductors are found on the diagonal, or below it, 
meaning, that they are undisturbed or lightly (negatively) affected 
by incoherent scattering effects. These are in fact stacked G:C molecules that 
conduct via delocalized states. %in the absence of the environment. 
%Mostly,  G:C sequences follow this trend. 
(ii) The poorest conductors tend to follow the scaling $G\propto \gamma_d^2$. 
These are mostly A:T sequences with charge transport proceeding via multi-site hopping
%which are ohmic conductors 
at finite $\gamma_d$.
(iii) While we can identify ballistic (green) and ohmic (black) conductors, most molecules 
display an in-between behavior, $G\propto \gamma_d^{\alpha}$ with $1\lesssim\alpha\lesssim2$ and  $0\lesssim\alpha\lesssim1$ 
at low and high electronic temperature,
respectively. 
These sequences conduct via an {\it intermediate}, coherent-incoherent mechanism,  which brings us to
one of the principal findings of our work:
%Within $1-5$ nanometer-long DNA molecules, 
The majority of 1-5 nanometer long DNA sequences conduct via a mixed quantum-classical mechanism. 
These molecules cannot be classified as tunneling barriers, ohmic conductors or  ballistic wires. 

% intermediate
Intermediate conduction mechanisms in DNA, distinct from both deep tunneling and multi-step hopping
were revealed in several other studies %including environmental  effects
performed e.g.  %with quantum mechanical tools, such as
with the surrogate Hamiltonian approach \cite{Berlininter}, 
the phenomenological Buttiker's probe method \cite{Ratner15}, and the
time-dependent stochastic Schr\"odinger equation   \cite{flickering,Beratan16}. 
Physically, an intermediate coherent-incoherent behavior results for example from polaron formation
involving charge delocalization over several bases \cite{Berlininter,Lewis}, or due to
the contribution of flickering resonances (achieved through conformational fluctuations) 
\cite{flickering}.

%=============================================
\subsection{Metal contact: Electronic temperature and hybridization energy}

\begin{figure}[htbp]
\includegraphics[width=15cm]{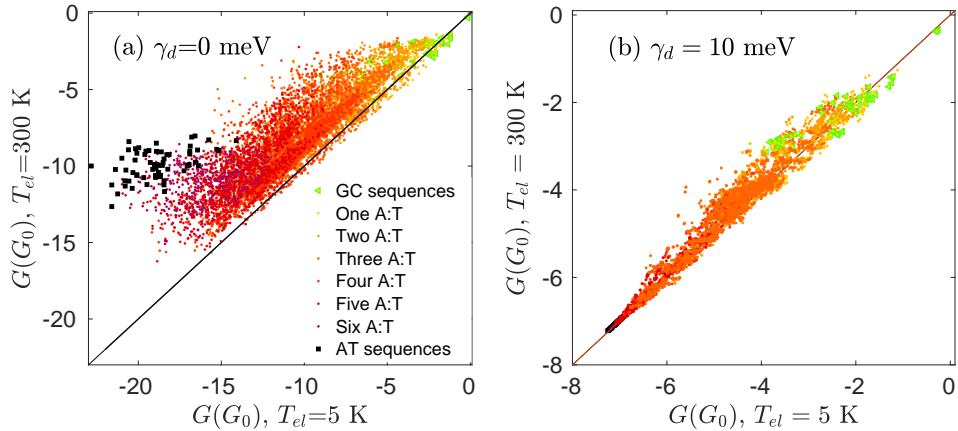} % scatterT1
\caption{Effect of the metal temperature $T_{el}$ on the conductance,
$n=7$, $\gamma_{L,R}=50$ meV,
(a) $\gamma_d$=0, (b) $\gamma_d$=10 meV.
}
\label{7bpT}
\end{figure}

Figure \ref{7bpT} demonstrates the subtle role of the electronic temperature in rigid and flexible molecules.
In the former, %($\gamma_d=0$), 
increasing the electronic temperature dramatically enhances the tunneling conductance, 
particularly, the conductance of sequences with A:T base pairs.
In contrast, scattering of carriers with environmental degrees of freedom results in their local 
equilibration at each site, washing out the the effect of the incoming charge distribution. 
A strong response of the conductance to electronic temperature thus indicates on 
the transition of the transport mechanism, from deep tunneling to resonant transmission,
rather than from tunneling to multi-site hopping conduction \cite{Kilgour1,Kim2}. % and more.

%---------------
% figure 6
%---------------

\begin{figure}[htbp]
\includegraphics[width=14cm]{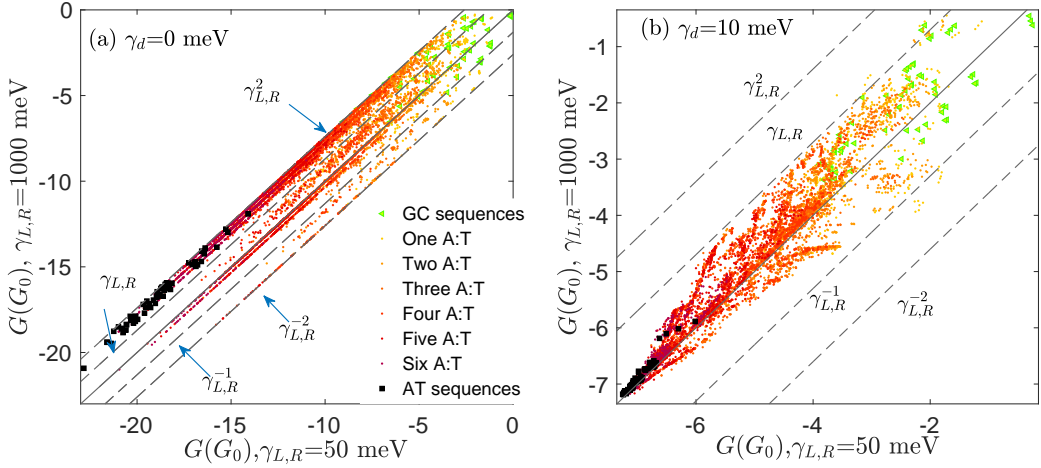} % scatter2
\caption{Effect of the metal-molecule coupling  $\gamma_{L,R}$ on the conductance
of $n=7$ bp DNA molecules, $T_{el}=5$ K.
(a) $\gamma_d=0$ meV, (b) $\gamma_d=10$ meV. 
The different scalings are marked by dashed lines.
}
\label{7bpglr}
\end{figure}

We argued above that most non-rigid sequences conduct via a mixed, coherent-incoherent mechanism.
It is not surprising therefore to find that the metal-molecule hybridization energy 
influences the conductance of these systems in a rather rich manner, as we show in Figure \ref{7bpglr}.
We find that the electrical conductance scales as $G\propto \gamma_{L,R}^{\alpha}$ 
where $-2\leq\alpha\leq2$  in rigid structures  %($\gamma_d=0$ in panel a)
but  $-1\leq\alpha\leq1$ at most when $\gamma_d\neq0$. 
A:T sequences in particular show a clear adjustment  from $\alpha\sim 2$
to $\alpha=0$ as we turn on the system-environment coupling, indicating on the conversion of transport mechanism
from tunneling to  multi-site hopping conduction, which is dominated by the bulk of the compound,
rather than interface effects.

%s limit, the electrode plays a negligible role, and the electronic conductance is dominated by the bulk.

\begin{figure}[htbp]
\includegraphics[width=14cm]{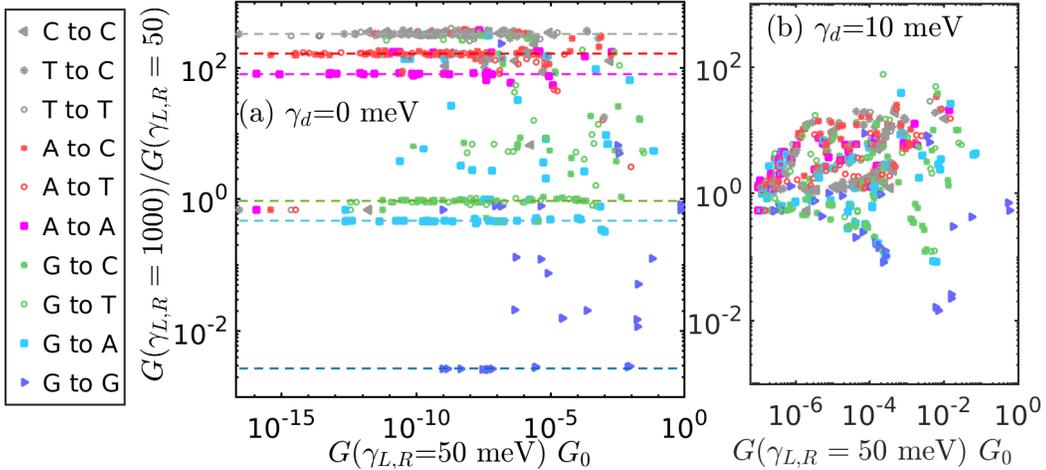}
\caption{Ratio of conductances at different values of $\gamma_{L,R}$ for an $n=5$ dsDNA.
Symbols correspond to different families of sequences, 
prepared based on the identity of the entry bases.
%depending on which bases connect to the source and drain electrodes.
%
For example,  gray triangles mark molecules with C bases at the boundaries,
the gray $*$ symbol indicates on molecules with T and C bases at the edges,
%Effect of metal-molecule coupling ($\gamma_{L,R}$) on the conductance
$T_{el}=5$ K, (a) $\gamma_d=0$ meV, (b) $\gamma_d=10$ meV.
Dashed lines highlight the scaling behavior of selected classes of molecules.
 }
\label{5bplr}
\end{figure}

Figure \ref{7bpglr} is involved, and it does not allow us to resolve underling principles.
What are the leading factors influencing the scaling of the conductance with $\gamma_{L,R}$?
Figure \ref{5bplr} resolves this question and brings to light the critical role of gateway sites.
For simplicity, we focus on a short DNA  with 5 bp.
We find that if both entry sites are the guanine base, 
the conductance is almost independent of $\gamma_{L,R}$ in  ballistic conductors such as
CGGGG. %%%
It diminishes with hybridization roughly as
$G\propto \gamma_{L,R}^{-2}$ when a barrier is formed, as in CGGAG or CTAAG \cite{nitzan}.
In contrast, when the gateway sites are situated far away from the Fermi energy, as in the case
of the C and T nucleotides,  increasing the broadening enhances the conductance by 
allowing better injection of carriers into the molecule.
Specifically, if both gateway stated are situated off-resonance, 
as in GAATC, the conductance scales as $G\propto \gamma_{L,R}^2$.
In mixed situations,
such as having G and T bases at the two boundaries in AGCCG, a nontrivial cancellation effect takes place and 
$\alpha\sim 0$. 
Finally, when carriers interact with the environment, the impact of the gateway groups lessens, 
becoming inconsequential for ohmic conductors.
In fact, when $\gamma_d\neq0$ the metal hybridization mostly affects
 {\it intermediate} coherent-incoherent conductors;
the poor conductors rely on multi-step hopping, which is rather insensitive to the interface,
and the exceptional band-like coherent conductors enjoy a good injection of charge, thus they
remain unaffected by the contact energy.

 %---------------
% figure 7
%---------------

%---------------
% figure 8
%---------------

\section{Conclusions}
\label{sec-summ}

Fundamental principles in DNA nanoelectronics are distilled here based on direct, all-inclusive simulations.
Experiments suggest that electron transfer in $1-10$ nm-long DNA is highly 
sensitive to subtle structural variations. %, see e.g. \cite{BartonNN11}. XXX
Our calculations support this observation. By tuning the effect of the nuclei on charge dynamics %(internal nuclei or environmental)
we find orders of magnitude enhancement of the electrical conductance, from the the deep tunneling coherent limit 
to the incoherent case.
While our calculations obviously suffer from major simplifications that limit the accuracy 
of calculated values, we bring forward underlying principles for DNA electronics at the nanoscale:
% Main results

(i) Mostly, natural DNA is a poor electronic material. %, supporting an electrical conductance $G<10^{-5}$ $G_0$.
%Considering all $n=3-7$-long sequences, most act as poor electronic conductors, with $G<10^{-5}$ $G_0$.
(ii) The conductance of rigid structures cannot serve as a proxy for the conductance of flexible molecules.
% since different factors affect their conductances. 
In particular, sequences that comparably conduct when frozen
may differ by up to 2 orders of magnitude when the molecular environment is allowed to
influence the transfer process.
(iii) Both the composition (content) and the structure (order) 
are important in determining the transport behavior of non-rigid molecules.
The conductance of sequences rich in G:C nucleotides is generally 
high compared to those containing predominantly A:T nucleotides.
DNA sequences with an island of A:T units show a lower conductance compared to sequences in which A:T 
base pairs are placed apart from each other.
(iv) Gateway states largely affect the conductance of rigid molecules. 
Moreover, the role of the metal-molecule contact is influential in many 
intermediate coherent-incoherent 
conductors.
(v) Lastly, the majority of DNA molecules examined here conduct via a mixed quantum-classical 
(coherent-incoherent)  mechanism. Only few, special sequences can be classified as  
tunneling barriers, ohmic conductors, or ballistic molecular 
wires. %We believe that longer molecules with 8-12 bp would bring in similar conclusions.
Our main finding is that quantum coherent effects are prevalent, and they play a central role in  
biological electron transport over the
distance of few nanometers. It is interesting to explore similar questions of
electron transfer through proteins \cite{Dies}.

In closing this computational search, which sequences come forward as good and robust conductors?
Not surprisingly, we find that conjugated G:C sequences with a {\it single} crossing between strands are superb
conductors with $G\sim 0.3-0.75$ $G_0$ across a wide range of parameters;
recall that the electrodes are connected to the 3' ends. 
The crossing between strands should minimally disturb the $\pi$ conjugation, 
thus the best e.g. $n=7$ sequence is CGGGGGG, 
while the next best sequences are CCGGGGG and CCCGGGG, and so on.  
These molecules are only lightly affected by the metal-molecule contact 
and the coupling to the surrounding environment. % and the electronic and nuclear temperature.
Other good and relatively robust, 
though less obvious sequences are those with an adenine at one end,
a segment of unperturbed G bases, and a single crossing between strands, for example,
CGGGGGA and TCGGGGG, coming up with  $G\sim 0.1$ $G_0$.
These molecules are robust against the nuclear environment, 
and strengthening the metal-molecule contact enhances their conductance.
According to our calculations, the best DNA electrical conductors support charge transport 
through delocalized states.
This  band-like coherent motion
is robust against environmental interactions. Nevertheless, it is important to remember that e.g.,
out of the 8192 different sequences with $n=7$ base-pairs, we identify here only very few ($\sim 5$)
excellent and robust conductors.

% defects
Our computational method suffers from several obvious shortcomings:
%(i) 
The electronic Hamiltonian is included at a coarse grained level, with each base represented by
a single electronic site using a fixed parametrization.
%
%(ii) 
A realistic molecule-electrode coupling model is missing in our treatment, and we
capture this coupling with a single parameter. % the hybridization energy.
%
%(iii) 
Furthermore, in all calculations we place the Fermi energy of the electrodes at the energy of the guanine base. 
The electronic structure could be improved by treating the contact atoms explicitly, and by
generating an electronic Hamiltonian for each sequence separately.
%
%(iv)
Another weakness of our framework is that the dynamics of the environment %and genuine electron scattering effects with the nuclei, are 
is not explicitly treated.
We encapsulate scattering effects of conducting charge carriers from different 
sources (low and high frequency phonon modes,
polarization effects, static and dynamical fluctuations) into a single, constant parameter
that dictates decoherence and energy relaxation.
One could improve our method from here by e.g. providing distinct scattering lifetimes on different sites,
and by averaging over a static disorder. 
Overall, state of the art QM/MD simulations are costly and impractical for our purposes,
while the low-level Landauer-B\"uttiker framework employed here
provides a meaningful starting point for performing large scale simulations beyond the coherent quantum limit.

Our minimal treatment of DNA nanoelectronics is not expected to be in quantitative 
agreement with experiments, but it is powerful enough to uncover fundamental
principles, most importantly, that the majority of inspected molecules display 
mixed coherent-incoherent charge transport characteristics.
We aspire this work to stimulate  %subsequently tested experimentall
experimental tests %to test our predictions,
as well as more precise calculations %over the electronic characteristics of DNA molecules,
that could guide the search for DNA molecules with desired electronic properties. 
Finally, chiral-induced spin selectivity through double helical DNA critically depends 
on the nucleobase fluctuations  \cite{Naaman1,Naaman2}. 
The present framework can be readily extended to explore spin-polarized charge current effects \cite{spin}.

%==========================================================
\begin{acknowledgments}
DS acknowledges support from an NSERC
Discovery Grant and the Canada Research Chair program.
The work of RK was supported by the CQIQC summer fellowship at
the University of Toronto and the University of Toronto Excellence Research Fund.\\
\end{acknowledgments}
%=========================

%===============================================

%=================================================

\begin{thebibliography}{0}
\bibitem{Consite}
In the parametrization of Ref. \cite{BerlinJacs}, on-site energies depend on the identity of neighboring bases.
This variation of energy from site to site, even in a uniform chain,
largely destroys coherent resonant dynamics, which can be revocered by the
occurances of ``flickering resonances", achieved through conformational fluctuations \cite{flickering}.
Here, we simplify this description and assign a single, averaged value
for on-site energies for each base, see Table \ref{table:en}.
As we argue in Ref. \cite{Korol1}, the combined effect of site energy variations with flickering resonances
is practically comparable to a model with fixed site energies suffering level broadeneing $\gamma_d$.
%
We also perfomed simulations directly with the parameter set of Ref. \cite{BerlinJacs},
without averaging over site energies.
These results are very similar to those reported in the main body of the paper,
only missing the coherent-molecular wire component---which should be revived by structural fluctuations.
\bibitem{gamma}
The matrices are given by
$(\hat \Gamma_{i\neq 1,N})_{p,q}=\gamma_d\delta_{p,i}\delta_{q,i}$,
$(\hat \Gamma_{1})_{p,q}=(\gamma_d+\gamma_L)\delta_{p,1}\delta_{q,1}$, and
$(\hat \Gamma_{N})_{p,q}=(\gamma_d+\gamma_R)\delta_{p,N}\delta_{q,N}$
%%
with the $i=1$ site coupled to the left metal, and site $N=2n$ coupled to the right lead.
\end{thebibliography}
\end{document}